# FMR Study of Co/Ti Bilayer Thin Films

M. Erkovan[1,2*], S. Tokdemir Öztürk[2], D. Taşkın Gazioğlu[2], R. Topkaya[2], O. Öztürk[2]

[1]*Gebze Institute of Technology, Department of Physics, Gebze, Kocaeli, Turkey.*
[2]*Sakarya University, Department of Metallurgy and Materials Engineering, Sakarya, 54687, Turkey*

**Abstract.** We focused on the interaction between two ferromagnetic cobalt layers through a non-magnetic titanium layer. The magnetic properties of the structure were characterized by ferromagnetic resonance technique (FMR). The data were collected as a function of non-magnetic titanium layer thickness. Co/Ti multilayer (Ti (50 Å)/Co(45 Å)/Ti(2-40 Å)/Co(40 Å)/Ti(100 Å))films were grown onto naturally oxidized p-type single crystal Si (100) substrate at UHV condition with magnetron sputtering system at room temperature. The thickness of Ti spacer layer ranges from 2 to 40 Å with 2 Å steps. We did not observe usual optic and acoustic modes; instead we had two broad overlapped peaks for the films ranged from 6 Å to 40 Å. One interesting result was the high anisotropic resonance field values for these films. Exchange coupling between ferromagnetic layers causes shift on resonance field values but these shifts in our samples were much larger than expected. This large anisotropic behavior is not clear at the moment. Our theoretical model was not able to determine a value for the exchange coupling parameter. One reason can be the close thickness values for Co sublayers. The other reason can be the Ti non-magnetic layer. If titanium did not grow layer by layer on cobalt, the cobalt ferromagnetic layers may behave as a single layer. As a result one cannot observe exchange interaction between ferromagnetic layers through non-magnetic spacer.

## Introduction

Magnetic multilayer films have interesting magnetic properties compared to single layer magnetic films such as giant magneto resistance effect (GMR),[1,2] tunneling magneto resistance effect (TMR).[3,4] Magnetic multilayer systems are composed of alternating ferromagnetic layers and non-magnetic layers. [5] If the non-magnetic material is metal, GMR effect may be observed. In order to observe TMR effect, the non-magnetic layer has to be insulating material. Another important magnetic property is observing oscillating interlayer exchange coupling which changes with magnetic and non-magnetic layer compositions and thicknesses.[6,7] The interlayer exchange coupling between two ferromagnetic layers is ferromagnetic (antiferromagnetic) when the magnetization vectors of magnetic layers are parallel ( anti-parallel). The magnetic multilayer films are grown by many different deposition techniques [8, 9]. But the films prepared with the magnetic sputtering-deposition process in Ultra High Vacuum (UHV) conditions have advantages such as growing conditions lead to the formation of more complete thin-film layers and consequently periodic behavior of exchange coupling parameter is clearly observable. The non-magnetic layer homogeneity is crucially important to identify the exchange coupling species. The quality of the homogeneity of films is very high [10] when the films are prepared with this technique. There are a lot of magnetic characterization techniques to investigate magnetic properties of magnetic multilayer films such as FMR, VSM, SQUID, MOKE [10, 11] etc. Especially ferromagnetic resonance is a very suitable technique to identify interaction species between ferromagnetic layers. Two different modes are generally observed in these structures depending on magnetic layer thickness, one of the modes is called optic mode and the other is called acoustic mode and their relative intensities and positions change with exchange interaction. If the exchange interaction is ferromagnetic (anti-ferromagnetic) the optic (acoustic) mode has lower resonance field value than the acoustic (optic) mode's value [12, 13]. There are some studies on Co/Ti multilayer structures in the literature [14-18]. One study focused on the structural transformation of cobalt from the polycrystalline to the soft nanocrystalline structure in Co/Ti multilayer system. FMR was used to determine the change in effective anisotropy field for the critical thickness of Cobalt [14]. P. Wu *et all.* investigated the magnetization and interface structures of the Co/Ti multilayer films. The saturation magnetization of the film is found to decrease linearly with $1/d_{Co}$ and also decrease with $1/d_{Ti}$ [15]. M. Schmidt

*et all*. used the Co/Ti films for pseudo spin valve structures with Cu layer [16]. These studies do not give any information about interlayer exchange interaction between two cobalt layers. On the other hand, Smardz used VSM technique in his study [17] to investigate the interlayer exchange interaction. Results showed some conditions to observe exchange coupling between two ferromagnetic cobalt layers. According to the author, Co sublayers are ferromagnetically coupled up to Ti spacer if thickness is about 19 Å. Furthermore, a weak antiferromagnetic coupling of the Co sublayers was observed for a Ti thickness range between 19 and 27 Å. In this study, Cobalt was chosen as ferromagnetic layer and Titanium was chosen for nonmagnetic transition metal as a spacer. The interlayer exchange coupling between ferromagnetic cobalt layers was investigated as a function of titanium thickness with ferromagnetic resonance (FMR) technique.

## Sample Preparation

Co/Ti multilayer (Ti (50 Å)/Co(45 Å)/Ti(2-40 Å)/Co(40 Å)/Ti(100 Å)) films were grown onto naturally oxidized p-type single crystal Si (100) substrate at UHV condition with magnetron sputtering system at room temperature. The thickness of Ti spacer layer ranges from 2 to 40 Å with 2 Å steps. RF power supply was used for sputtering the cobalt magnetic layers, and DC power supply was used for sputtering the titanium layers. Both cobalt and titanium deposition ratio were determined by X Ray Photoelectron Spectroscopy (XPS) and Quartz Crystal Monitoring (QCM). All details about deposition ratio determination with XPS are given in reference 18. Before the substrates were loaded to the ultra high vacuum chamber, they were cleaned with ethanol and methanol for ten minutes by ultrasonic cleaner. In order to remove surface roughness, the substrates were heated at UHV conditions by PBN heater and hold at 600 °C for 20 minutes. Although the system base pressure was about $5 \times 10^{-8}$ mbar, the deposition pressure was about $1.2$-$1.5 \times 10^{-3}$ mbar. The water-cooled target with 3 in. in diameter provides a good homogeneity in thickness. High purity argon gas (6N) was used for sputtering.

From earlier experiences [13] it was found that measurable exchange coupling could be observed between ferromagnetic layers through a non magnetic spacer when the thicknesses of bottom and upper ferromagnetic layers were different but close to each other. For that purpose the thicknesses of bottom and upper Co layers were chosen as 45 Å and 40 Å, respectively. The buffer titanium layer was 50 Å to remove substrate effects and the cap layer titanium was 100 Å to protect films from atmospheric effects.

## Experimental Results

A Bruker EMX model X-band electron spin resonance spectrometer was used for FMR measurements. The microwave frequency was 9.5 GHz. Two measurements were taken as a function of the angle of the external dc field with respect to the film normal at room temperature. The sample sketch, relative orientation of the equilibrium magnetization vector **M**, the applied dc magnetic field vector **H**, and the experimental coordinate system are shown in Fig. 1(a). When the magnetic field is parallel to the plane of the film we call this position as in plane geometry (IPG) and when the magnetic field is perpendicular to the plane of the film we call it out of plane geometry (OPG). The picture of the prepared trilayer structure is shown in Fig. 1(b). The magnetic field component of microwave is always kept perpendicular to the dc field during the sample rotation. The applied microwave field always remains in sample plane for conventional geometry and power is kept small enough to avoid saturation, as well. A small modulation field of 100 kHz was applied in parallel to the dc magnetic field in order to record the field derivative of absorption power.

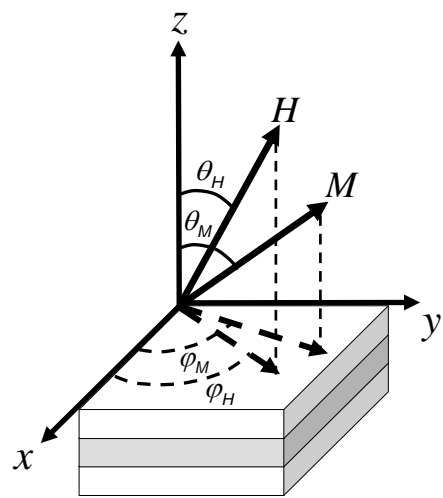

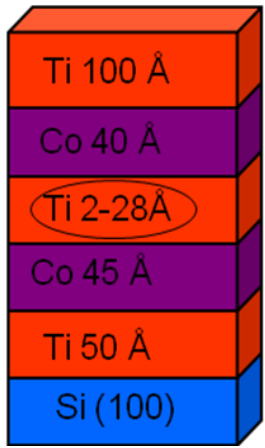

**FIGURE 1.** ( a) Representative picture of Ti (50 Å)/ Co (45 Å)/ Ti (2-28 Å)/ Co (40 Å)/ Ti (100 Å) multilayer structures. (b) Relative orientations of the external dc magnetic field and magnetization vectors with respect to sample plane.

Figure 2 shows the FMR spectra for different Co thickness when the external magnetic field is parallel to the film plane. Only one FMR mode with narrow line width was observed for the thinnest spacers up to 6 Å. The signals get broader as the spacer thickness increases and single FMR signal becomes doublet. Especially 30 Å and 32 Å films show this feature clearly. Magnetic resonance positions are nearly the same for all films; however resonance positions shift to the lower magnetic field values for 10 Å, 20 Å, and 30 Å.

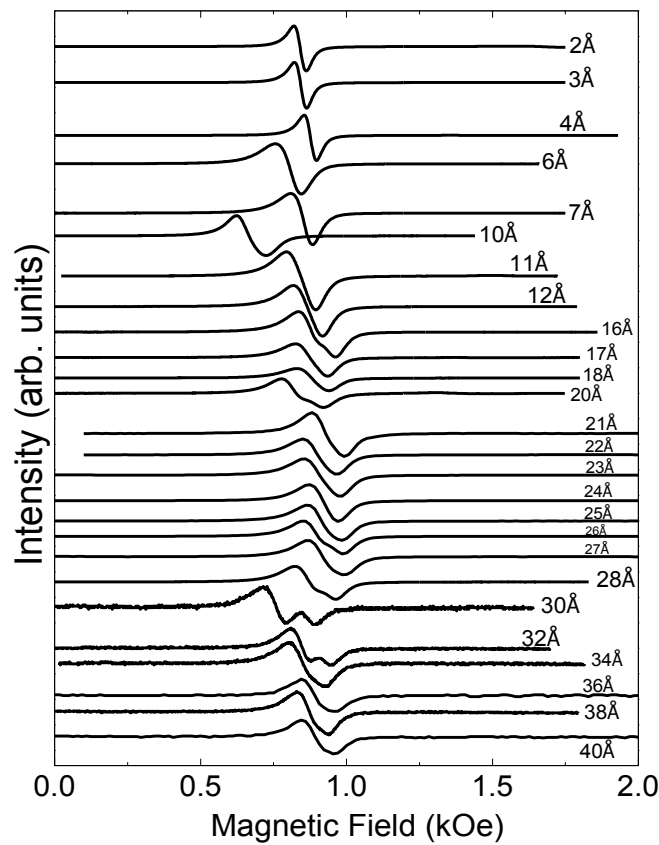

**FIGURE 2.** Experimental FMR spectra of the cobalt single layer films with different spacer thickness when the external field is parallel to the sample plane.

The FMR spectra for the perpendicular position are shown in Figure 3 (a). A similar situation was observed for this case too. For thin spacers only one narrow FMR mode was observed up to 6 Å. As spacer (Ti) increases two not well-resolved FMR modes were observed. However in perpendicular position, the spacer thickness plays an important role in resonance positions of these modes. The dependence of the resonance field on the spacer thickness is much stronger for the optical mode. Figure 3 (b) shows how sensitive the resonance positions are to the thickness of the spacer. One can see an oscillation behavior. Roughly peaks occur for 10 Å, 20 Å, and 30 Å films. The distances between two modes do not depend on the spacer thickness.

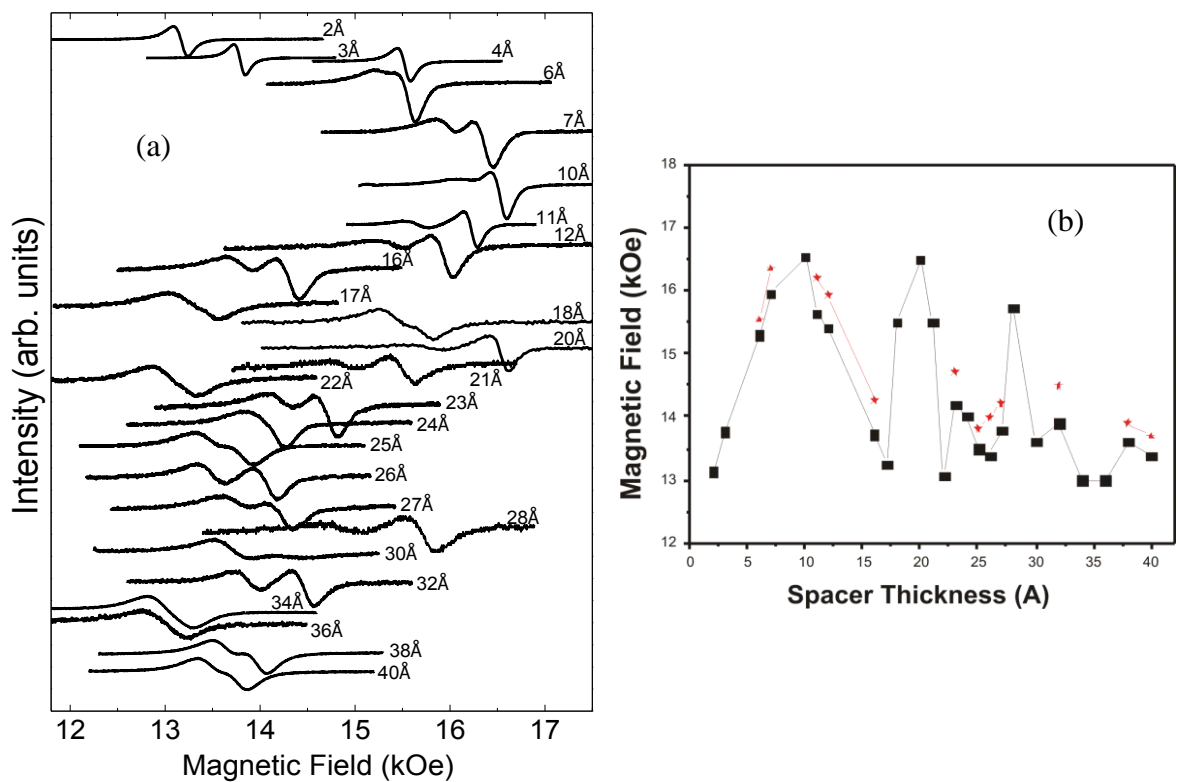

FIGURE 3. Experimental FMR spectra (a) and resonance positions (b) of the cobalt single layer films with different spacer thickness when the external field is perpendicular to the sample plane.

## Simulation of The FMR Data

The collected FMR data was tried to be analyzed with the theoretical model which was explained in details in reference 12. The model is suitable for a system consisting of N magnetic layers with saturation magnetization Ms and layer thickness. However the computer program developed for this purpose was not able to fit the experimental spectra to the theoretical model. The position of the resonance peaks for each ultrathin Co(45 Å)Ti (2-28 Å) Co(40 Å) trilayers were not able to be determined by any suitable saturation magnetization $Ms$, effective anisotropy and exchange coupling of the magnetic layers.

The doublet peaks for particular Ti thickness were expected to represent the optical and acoustic modes. We know that if ferromagnetic layers are magnetically equivalent, a single resonance peak is observed due to simultaneous excitations of precession of magnetization in all layers. On the other hand, if layers are magnetically nonequivalent two resonance modes are observed in FMR curves for OPG case as a result of exchange coupling of magnetically nonequivalent neighboring layers. If magnetic properties of different layers are very close to each other, two modes come close to each other and additionally if the damping parameter is relatively larger, these two peaks overlap and give a distorted single line. Choice of 40 Å and 45 Å Co layers could be the case for inaccuracy of exchange parameter.

However in this study, the anisotropy of the resonance field values as a function of Ti thickness is very large and periodic (nearly 10 Å). The reason for this large shift is not obvious. Even though there is small exchange interaction between different layers; this cannot be responsible for such a large shift.

## Conclusion

There is a long-range oscillatory indirect magnetic exchange coupling between two ferromagnetic layers separated by thin layers of the nonmagnetic transition metals. Parkin [22] showed that this spectacular phenomenon occurs with almost any metal as the spacer material. As a result of this indirect exchange interaction two different resonance modes (optical and acoustic) are observed on the FMR spectra. Their relative resonance field values and intensities help to characterize the magnetic interaction between ferromagnetic layers. In this study, we did not observe two well resolved modes for all samples; instead we had two broad overlapped peaks. Unfortunately, these broad peaks were not able to be identified by the theoretical model which is suitable for thin multilayer systems. Smardz [19,20,22] observed that Co sublayers are very weakly exchange coupled or decoupled for dTi > 27 Å in the previous hysteresis measurements for 170 Å Co-dTi-170 Å Co trilayer. Also he didn't find any indication for the antiferromagnetic coupling between Co sublayers. In our experiment, FMR spectra support that Co sublayers are very weakly exchange coupled for dTi > 6 Å. One interesting thing is that the thickness of Ti spacer does not affect the magnitude of the exchange coupling but rather affects the magnetic resonance positions in a periodic way.

There are two possible reasons for not clearly observing exchange coupling interaction between ferromagnetic layers. The first reason is the choice of two magnetic layer thicknesses. In our experiment thickness differences between Co sublayers were only 5 Å. It seems this choice may not be the right one. However, pure Co 40 Å and Co 45 Å films gave resonance peaks at different magnetic field positions, so we do not think that sublayers thicknesses is the issue here. The second reason may be the choice of non-magnetic spacer. If titanium did not grow layer by layer it did not behave as spacer between magnetic layers. As a result, two Co magnetic layers behaved such as one layer and simultaneous excitations of precession of magnetization occurred in all layers.

## Acknowledgments

This work was partly supported by TUBITAK (Project No:TBAG-106T576) and by State Planning Organization of Turkey (DPT-Project No:2007K120900). We gratefully acknowledge that all samples used in this study were grown at the Nanotechnology Center of Gebze Institute of Technology.